\documentclass[sigplan,nonacm]{acmart}

\usepackage{graphicx}
\usepackage{subcaption}
\usepackage{booktabs}

\begin{document}
\title{Exploiting Multicast for Accelerating Collective Communication}
\author{Chao Xu, Xu Zhang, Zihang Luo, Yuyan Wu, Guoxin Qian, Yufeng Yao,\and CHIHYUNG WANG, Jingbin Zhou}

\affiliation{
  \institution{Huawei}
  \city{Shanghai}
  \country{China}
}

\begin{abstract}
Reducing collective communication latency is a critical goal for large model training and inference in both academia and industry. Many-to-many communications, such as AllGather and AlltoAll (dispatch), are core components of modern parallelization strategies. State-of-the-art implementations of these communications rely on unicast-based writes and transmit duplicate copies of the same data across physical links for multiple receivers. This redundant transmission congests network bottlenecks and degrades end-to-end latency. We present MultiWrite, a novel many-to-many transmission semantic that eliminates redundant packets to directly reduce operator latency. MultiWrite adopts multicast principles while addressing critical limitations of traditional multicast for AI workloads. These limitations include heavy management plane overhead and ecosystem compatibility issues. We implement MultiWrite on Ascend NPUs. Long-term stress tests demonstrate that our MultiWrite-based operators achieve up to 33\% latency reduction on commercially deployed devices.
\end{abstract}


\maketitle

\section{Introduction}

As large model parameter scales grow from hundreds of billions to trillions, distributed training and inference have become the foundational paradigm in AI (Artificial Intelligence) \cite{narayanan2021efficient, fedus2022switch, zheng2022alpa}. To support ultra-large models with limited hardware resources, the industry widely adopts data \cite{li2020pytorch, sergeev2018horovod}, tensor \cite{shoeybi2019megatron}, pipeline \cite{huang2019gpipe}, and expert \cite{fedus2022switch, lepikhin2020gshard} parallelism to decompose computation. However, these strategies introduce frequent and intensive collective communication \cite{nvidia2026ncclcollectives} between accelerators, such as GPUs and NPUs \cite{xu2025hardware, lee2021architecture}. Operations like AllReduce, AllGather, and AlltoAll now directly determine the scalability and end-to-end efficiency of distributed systems. Empirically, communication typically consumes 20-30\% of runtime in large-scale training tasks, evolving from a supporting role to a core performance determinant.

Among all collective communications, many-to-many patterns, such as AllGather and AlltoAll (dispatch), are particularly critical. In such transmissions, each node transmit duplicate copies of the same data to multiple receivers. When these copies traverse bottleneck links, communication latency grows significantly, and this in turn hurts training throughput and inference performance. Two trends have amplified this issue. First, communication domains are expanding rapidly\cite{yang2025hybridep}. The large EP technology introduced after DeepSeek V3 \cite{deepseekai2025deepseekv3technicalreport} allows AlltoAll domains to easily reach 64, 128, or even 256 nodes, creating more potential bottlenecks. Second, leading service providers now target TPOT (Time Per Output Token) below 5ms \cite{vllm2024perf}. At this stringent requirement, even minor degradation in communication latency, which may account for 30\% of end-to-end time \cite{he2026efficient}, can derail performance targets.

Unfortunately, current approaches to addressing communication issues remain reactive and fragmented.
Communication performance degradation is typically detected after training or inference jobs are running in production environments, and manifests as significant throughput losses or degraded end-to-end latency.
Engineers then rely on vendor-specific custom diagnostic tools to inspect hundreds to thousands of devices across the cluster in order to pinpoint the exact problematic ports, links, or routing configurations. The root cause analysis may last from several days to several months \cite{gu2026ccl, lin2025understanding}.
Even after root cause identification, fixes are ad-hoc: switching AlltoAll algorithms, adjusting routing weights, or adding communication-computation overlap. These tricks only work for specific hardware, parallel strategies, or model architectures. When workloads or clusters change, optimizations often break, forcing the entire cycle to repeat. This post-hoc firefighting paradigm is incompatible with the rapid iteration pace of AI.

To address the issue of communication latency degradation caused by the transmission of redundant data on bottleneck links, we present the MultiWrite semantic based on our long-term experience and practice. The core idea behind MultiWrite is straightforward. Since the problem arises from redundant data traversing bottleneck links, and bottleneck links are inevitable in any network topology, we eliminate redundant data injection into the network entirely. Specifically, in many-to-many communication, only one copy of the same data will be sent to the downstream, regardless of how many destination nodes it needs to reach. Data is then replicated and forwarded at appropriate intermediate nodes to satisfy the packet reception requirements of all receivers. This technique is essentially a specialized form of multicast. Applying multicast to collective communication is not a novel concept, and prior works have proposed this idea. However, the industry has been reluctant to adopt this technology for several reasons, which are precisely the challenges we address in this work.

The first challenge is the concern of whether multicast can really reduce latency. While multicast undoubtedly reduces bandwidth overhead on source egress links, but this does not necessarily mean multicast can reduce communication latency. Since collective communication transmissions typically traverse multiple links, bottleneck links may not be limited to the source node egress.
Furthermore, traditional multicast implies that packet switching devices must support packet replication and forwarding, a capability that is absent from most current AI clusters, especially in scale-up domains.

Second, the traditional multicast capability does not match the actual requirements of AI training and inference scenarios. As is well known, since DeepSeek v3, MoE (Mixture of Experts) and its corresponding large EP (Expert Parallelism) technology have become the de facto standard in large language models. The communication in MoE involves two all-to-all operations, also known as dispatch and combine. During the dispatch process, a single token of data needs to be sent to the top-$k$ experts, where $k$ is typically 8 or possibly more. The target experts are determined by the gate network. In this process, the destination nodes of two tokens emitted from the same source may be different, so these two tokens cannot be transmitted through the same multicast group. To meet the requirements of typical scenarios, the number of multicast groups would explode to a level of $10^{9}$, which is simply unacceptable.

Third, traditional multicast is incompatible with existing training and inference ecosystems. Under the hood, collective communication operators are implemented by decomposing them into multiple individual transfers, typically using unilateral write semantics. Traditional multicast requires additional operations for multicast group creation, management, and membership changes, as exemplified by IB (InfiniBand) UD (Unreliable Datagram) mode multicast implementation\cite{ibta2024infiniband}. These operations have no place in current collective communication stacks. Integrating them would necessitate a full framework redesign, which is prohibitively expensive.

MultiWrite addresses all three challenges. First, we identify and rigorously analyze non-niche scenarios where multicast-like techniques deliver measurable end-to-end latency gains, not just bandwidth reductions. Crucially, MultiWrite requires no changes to switches, which eliminates the primary obstacle to practical deployment. Second, we redesign multicast to carry destination information in packets rather than relying on pre-provisioned multicast groups, avoiding the group explosion problem in MoE scenarios. Third, we implement MultiWrite as an enhanced write semantic, not a separate module, preserving existing framework interfaces and usage patterns.

We implement MultiWrite on Ascend NPUs and build optimized AllGather and AlltoAll operators on top of it. Long-term stress tests on commercially Ascend devices show 12\%-33\% latency reduction for these many-to-many operations.
To summarize, this paper makes the following contributions:

\begin{itemize}
\item From a practical AI infrastructure perspective, we identify and characterize scenarios where multicast-like techniques deliver latency benefits, addressing the long-standing benefit-cost concerns that have hindered industrial adoption.

\item We design and implement the hardware-agnostic MultiWrite semantic, which solves the key limitations of traditional multicast while maintaining full compatibility with existing frameworks. Our implementation is easily portable to other accelerator platforms.

\end{itemize}

\section{Background and Motivation}

\subsection {Collective communication}
Collective communication denotes a suite of synchronization and data exchange operations involving all nodes in a distributed computing cluster \cite{nvidia2026ncclguide}, forming the backbone of large-scale AI training and inference systems. Core communications include point-to-point operations and collective operations, such as AllGather and AlltoAll. AllGather collects fragmented data partitions from all nodes to construct a complete dataset, a critical building block for tensor parallelism in transformer-based models. Strictly speaking, canonical AlltoAll enables pairwise data exchange between every node pair, functioning as a distributed matrix transpose where each data partition is sent to exactly one target node \cite{nvidia2026ncclcollectives}. However, with the emergence of sparse models like DeepSeek's MoE architecture, a variant of AlltoAll has become prevalent. In this variant, a single data partition may be dispatched to multiple nodes, and this pattern is widely referred as dispatch and combine\cite{deepseekai2025deepseekv3technicalreport}. Unless otherwise specified, all references to AlltoAll in this paper refer to this dispatch-style variant.

\subsection {Multicast for collective communication}
In AI training and inference frameworks, major vendors have developed their own collective communication libraries for upper-layer applications, such as NVIDIA's NCCL \cite{nvidia2026nccl} and Huawei's HCCL \cite{huawei2026hccl}. These libraries share a common implementation paradigm, building complex collective operations from numerous one-to-one unicast transactions. For many-to-many collective communications, this design inherently injects redundant data copies into the network, as the same data segment must be repeatedly transmitted to multiple target nodes.

Naturally, multicast has been proposed to eliminate redundant data in such transmissions. However, existing multicast-based solutions primarily demonstrate bandwidth load reduction at source node egress links, without clarifying the specific scenarios where bandwidth savings translate to tangible communication latency reduction\cite{huang2023mc, li2024cepheus}. These two effects are not always equivalent, as we will demonstrate in Section \ref {sec: motivation1}. Furthermore, these prior works either rely on specialized fixed hardware or require switches with advanced programmable capabilities. Such hardware dependencies further hinder the practical deployment of multicast in real-world AI training and inference clusters.

\subsection {Motivation}
While the idea of applying multicast to collective communication has long been proposed, we as AI infrastructure vendors have identified several unresolved questions in existing studies, which motivates the design of our novel MultiWrite semantic.

\subsubsection {Bandwidth reduction does not equate to latency reduction}
\label{sec: motivation1}
\begin{figure}[htbp]
	\centering
	\includegraphics[width=1.0\linewidth]{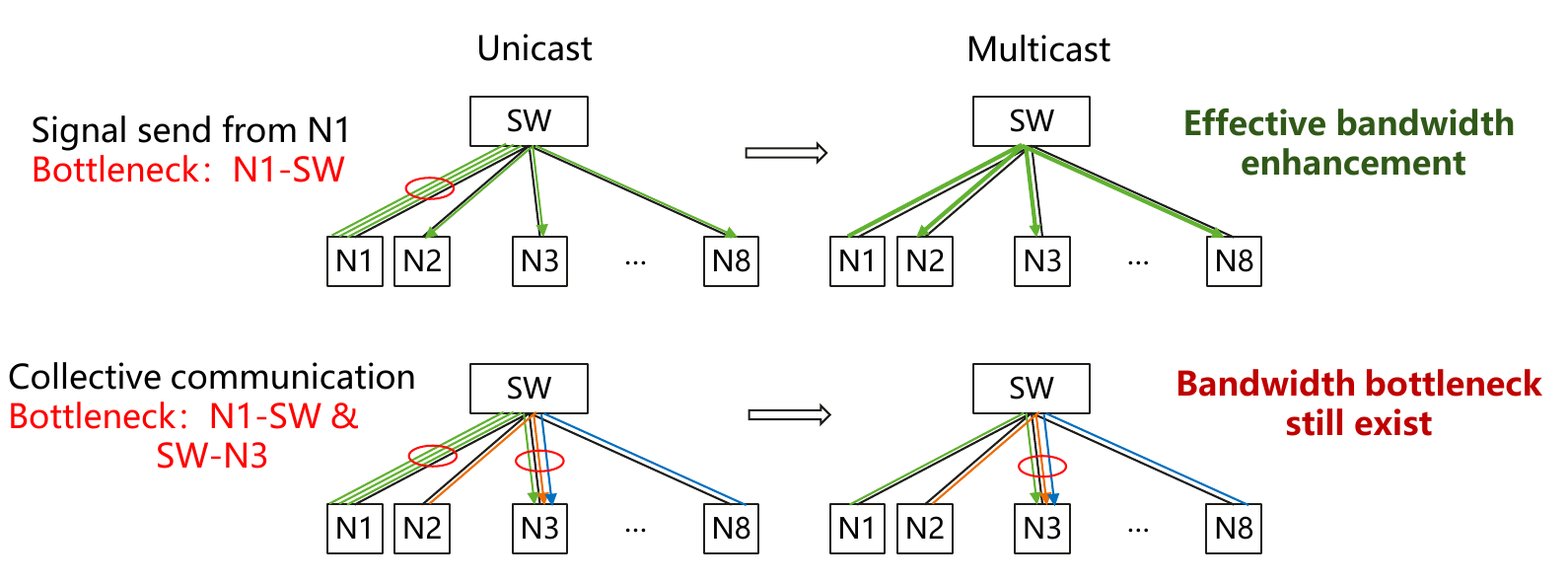}
	\caption{In typical CLOS topologies, multicast improves bandwidth for single senders but cannot enhance effective end-to-end bandwidth in collective communication, where bottlenecks exist on both uplink and downlink.
	}
	\label{fig:startopo}
\end{figure}

Consider the topology illustrated in Figure \ref{fig:startopo}, where eight NPU nodes are connected to a single switch. This is a classical CLOS topology in AI clusters. Assume that the eight nodes perform the AllGather operation. We first analyze the AllGather operation without multicast support, taking N1 as a representative source node. To complete AllGather, N1 must transmit its local data partition \texttt{Data1} seven times across the network. This results in seven redundant copies of \texttt{Data1} traversing the uplink between N1 and the switch. Denoting the bandwidth of each link as $W$, the effective bandwidth available for data transmission from N1 to any destination node is $W/7$, severely limiting communication efficiency due to redundant data injection.

We then examine the scenario with switch-level multicast support. N1 sends a single copy of \texttt{Data1} to the switch, which then replicates it into seven copies and forwards them to seven targets. In this case, both the uplink and the downlink carry only non-redundant data. This easily creates the misconception that the effective bandwidth per source-destination pair increases to
$W$, and the communication latency decreases to one third of the unicast baseline. Unfortunately, this conclusion does not hold in collective communication. AllGather means every node acts as both a source and a destination. For example, the downlink from the switch to N1 must carry other seven data partitions \texttt{Data1}, \texttt{Data2} $\sim$ \texttt{Data8}. The bandwidth load on this downlink thus remains unchanged with or without multicast. Multicast only reduces the uplink bandwidth load in such classical CLOS topology, while having no impact on the downlink load. As a result, the overall latency of AllGather stays the same, despite the fact that the uplink bandwidth load from each node to the switch is significantly reduced.

\subsubsection {Multicast group explosion in MoE scenarios}
Since the release of DeepSeek V3.1, the Mixture of Experts architecture has become one of the de facto standards for large language models. It has also emerged as a fundamental benchmark for evaluating AI infrastructure hardware and software stacks across vendors. However, a fundamental limitation arises when applying multicast to MoE scenarios. The number of required multicast groups becomes prohibitively large, which makes traditional multicast approaches impractical for real-world deployment.

There are two AlltoAll operations in MoE scenarios, combine and dispatch. The combine phase returns computation results from experts back to their original senders. It involves no redundant data transmission and thus does not benefit from multicast. In contrast, the dispatch phase routes each token to the top-$k$ selected experts, which inherently injects redundant data into the network
\footnote{When multiple tokens are destined for the same expert, two transmission modes are typically used. One aggregates these tokens into a data unit and triggers one transmission operation. The other sends tokens one by one. Both are in practical use with distinct advantages, and only the token-by-token mode is compatible with multicast deployment.}.
Traditional multicast implementations require pre-established multicast groups, each identified by a unique multicast group ID. During data transmission, packets carry the corresponding group ID in their headers. Multicast-enabled switches parse this field and replicate packets to all members of the specified group. Consider a typical MoE scenario with EP = 64 and Top-8 routing. Since the destination experts for each token are determined dynamically at runtime, the worst-case number of distinct destination node combinations is given by the combinatorial number $C(64,8)$. This value exceeds 4.4 billion, which is clearly impractical for any real-world deployment. We have also explored an alternative approach that uses a limited number of multicast groups with dynamic node join and leave operations. However, these control-plane operations typically introduce tens of microseconds of latency, which is incompatible with the strict low-latency requirements of the data plane.

\subsubsection {Incompatibility between multicast operations and communication operator frameworks}

Even if we deliberately narrow the scope of multicast application to optimize only operators with static destination sets such as AllGather, significant compatibility challenges remain. The core issue is that traditional multicast architectures operate on entirely separate control plane and data plane stacks, which are fundamentally misaligned with the design principles of existing collective communication operator frameworks.

Let us take the multicast operations defined in InfiniBand Specification Release 1.8 as an example. The multicast control plane includes a comprehensive set of management operations such as multicast group creation, join, leave, deletion and group pruning. On the data plane, multicast requires additional operations to attach a queue pair to a multicast group and detach a queue pair from a multicast group. Furthermore, InfiniBand networks incorporate two dedicated hardware components, the Subnet Manager and Subnet Administration, to handle multicast group management and provide group information storage and query services respectively. This entire multicast infrastructure is excessively heavyweight, and no provisions exist in current collective communication operator frameworks to accommodate these operations and components.
Most of communication operators are implemented using simple \texttt{write/read} operations.
Achieving full compatibility would require fundamental architectural changes to the operator framework, which represents a substantial engineering effort.
\section{Beneficial Scenarios for Multicast}

In this section, we present real-world scenarios where multicast adoption delivers measurable latency improvements for collective communication operators, rather than merely reducing bandwidth overhead. We present a detailed analysis of two representative scenarios: 8P node interconnect (\ref{sec:8pfullmesh}), and MoE with server bandwidth convergence (\ref{sec:bandconvergence}). Then, we conclude the first principles of multicast performance benefits based on these scenarios(\ref{sec:Principles}).

\begin{figure*}[htbp]
	\centering
	\includegraphics[width=0.8\linewidth]{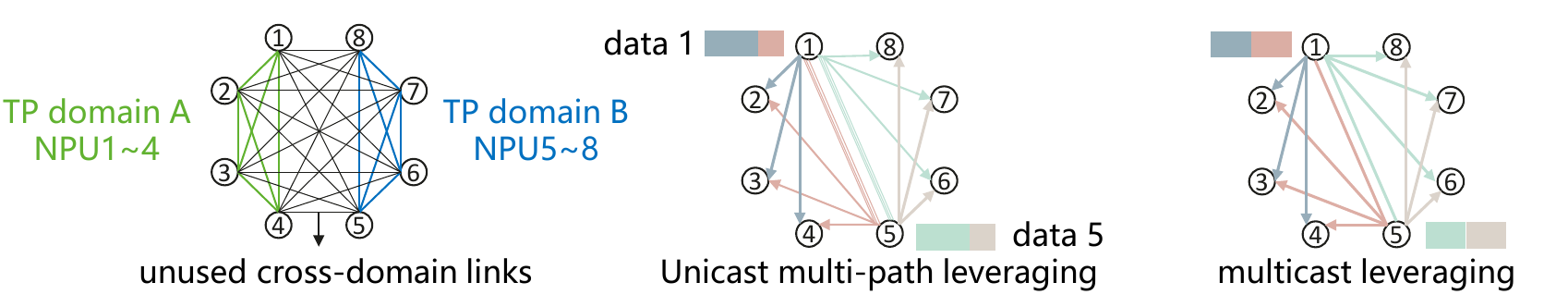}
	\caption{Left: Full-Mesh topology with two TP domains, leaving cross-domain links unused. Middle: Unicast multi-path leveraging utilizes cross-domain links but introduces redundant transmissions. Right: Multicast scheme exploits cross-domain links without redundancy, achieving higher effective bandwidth.
	}
	\label{fig:8ptopo}
\end{figure*}

\subsection {Scenario: Full-Mesh Topology}
\label{sec:8pfullmesh}
\textbf{Scenario Description:}
As illustrated in Figure \ref{fig:8ptopo},
Eight NPUs form a full mesh network. Each NPU uses a dedicated port and direct link to communicate with every other NPU, with no intermediate switches involved. In this topology, when 8 nodes are divided into two separate collective communication domains, multicast can reduce the latency of collective communication operations.

\noindent \textbf{Benefit Derivation:}
We illustrate this with the most typical TP = 4 configuration. NPU 1 to NPU 4 form one communication domain and perform an AllGather operation. NPU 5 to NPU 8 form the other domain. Let the bandwidth of each link be $w$, and each NPU holds a data fragment of size $s$. If we ignore other overheads such as fixed hardware startup latency, the AllGather transmission latency in each domain is $s/w$.

Clearly, a significant portion of links remains unused in the above case, as indicated by the black links in the figure. Cross-domain multi-path leveraging provides a straightforward method to improve end-to-end effective bandwidth. Each source-destination pair can use not only its direct link but also the unused cross-domain path to deliver data fragment. For clarity, we present a method called paired relaying. [NPU 1, NPU 5], ... , and [NPU 4, NPU 8] form respective pairs. Each NPU in a pair help its partner relaying data to all other NPUs in its partner's communication domain. For example, NPU 5 forwards data from NPU 1 to NPU 2, 3 and 4. At the same time, NPU 1 forwards data from NPU 5 to NPU 6, 7 and 8.

If multi-path leveraging is implemented via unicast, three identical copies of the relayed data will traverse the link from NPU 1 to NPU 5. Thus, the effective bandwidth of path NPU1-NPU5-NPU2 $w/3$. Combined with the direct link, the effective end-to-end bandwidth from NPU1 to NPU2 becomes $4w/3$, and the transmission latency reduces to $3s/(4w)$.

In contrast, if multi-path leveraging is implemented via multicast, NPU1 only send one copy of data to NPU5. NPU5 replicates the data into three copies and forwards them to NPU2, NPU3 and NPU4 respectively. Thus, the effective bandwidth of path NPU1-NPU5-NPU2 is $w$. Combined with the direct link, the effective end-to-end bandwidth reaches $2w$, and the transmission latency reduces to $s/(2w)$. This represents a 50\% latency reduction compared to the baseline (without multi-path leveraging), and a 33\% reduction compared to the unicast-based multi-path leveraging.

The paired relaying approach used in the above example is one of many possible strategies. Another approach, which we call full multi-path relaying, allows each NPU to leverage all NPUs in the opposite communication domain for data relay. Under full multi-path relaying strategies, multicast-based leveraging can also achieve a latency reduction of at least 16\%. Detailed derivation is omitted here due to space constraints.

\noindent \textbf{Real-world Cases:}
This above scenario is extremely common in real AI training and inference workloads. For example, the full-mesh topology is adopted in both Huawei Ascend 910B and 950 accelerators. On such hardware platforms, we have observed numerous real-world AI workloads that exhibit the scenario derived above.

\begin{itemize}
\item Tensor parallelism with degree 4.
\item Classifier-free guidance parallelism with degree 2 in multimodal generation \cite{ho2022classifier, chen2026parallel}.
\item Multi-tenant scenarios in cloud environments.
\end{itemize}

\begin{figure}[htbp]
	\centering
	\includegraphics[width=1.0\linewidth]{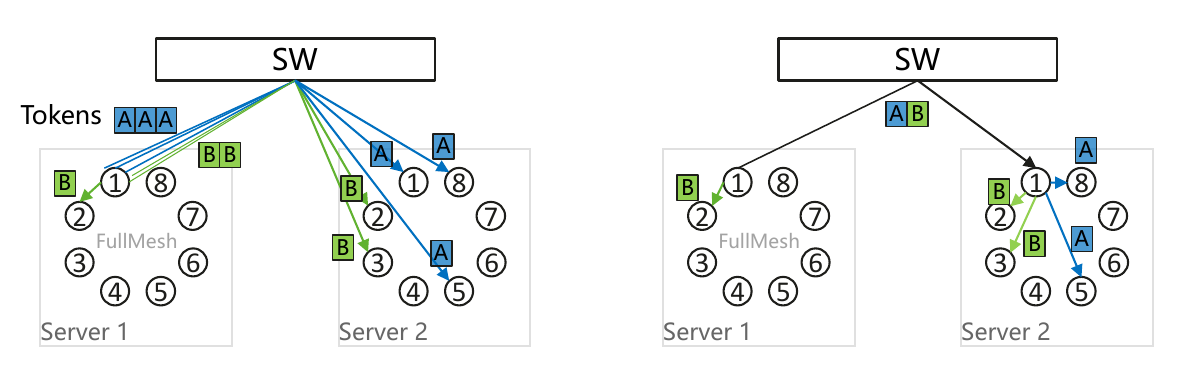}
	\caption{Multicast removes redundant data transfers on bandwidth-constrained inter-server links, thus improving effective communication efficiency.}
	\label{fig:crossserver}
\end{figure}

\subsection {Scenario: Oversubscribed Inter-Server Bandwidth}
\label{sec:bandconvergence}
\textbf{Scenario Description:}
As illustrated in Figure \ref{fig:crossserver}, each server houses multiple NPUs (e.g., eight devices per server) interconnected with high intra-server bandwidth, while servers are linked through a CLOS network fabric. The cross-server bandwidth is intentionally oversubscribed relative to the intra-server bandwidth, so the available bandwidth from an NPU to another NPU inside the same server is markedly higher than that to an NPU in a different server.

\noindent \textbf{Benefit Derivation:}
When a collective communication group grows large enough to span multiple servers, multicast-like techniques can exploit this asymmetry to reduce traffic. A prominent case is Expert Parallelism (EP) in Mixture-of-Experts, where the EP group size often exceeds 8, such as 64 and 128. Consider a scenario where NPU0 in Server0 dispatch a token to eight experts, four residing in Server0 and four in Server1. The overall dispatch latency is dictated by the slower of the intra-server and inter-server transfers, which is typically the inter-server transfer because of the lower cross-server bandwidth.

With a multicast approach, the intra-server communication proceeds as before, but instead of sending several copies across servers, NPU0 sends only a single copy to the same-index NPU on the remote server (e.g., NPU0 in Server1). Then, the same-index NPU replicates the token and forwards it to the target NPUs inside the server. Although the random placement of experts prevents us from providing specific benefit figures, the benefit qualitatively increases with larger MoE batch sizes and with a higher oversubscription ratio between intra- and inter-server bandwidth.

\noindent \textbf{Real-world Cases:}
Such oversubscribed topologies are common in real-world systems because providing full bisection bandwidth between servers is costly. Many users purchase individual servers or even discrete NPU cards and deploy them according to workload-specific traffic patterns to maximize cost performance. Representative devices such as Ascend 910B and 950 series are widely deployed with a noticeable oversubscription between intra-server and inter-server bandwidth.

\subsection {First Principles of Multicast Benefits}
\label{sec:Principles}
Compared with the scenarios presented in Section \ref{sec: motivation1}, the fundamental reason why multicast achieves latency benefits in the above two scenarios lies in the bandwidth reduction on bottleneck links.

In the first scenario, multiple flows share the same cross-domain physical link, making the shared path bottleneck. In the second scenario, inter-server link bandwidth is lower than intra-server bandwidth. By placing data replication at remote NPU nodes rather than conventional switches, the bandwidth overhead on the whole inter-server bottleneck links is effectively reduced. Although this approach increases the load on intra-server links and turns them into new bottlenecks, the achievable bandwidth of these new bottlenecks remains higher than that of the original inter-server links.

Therefore, AI infrastructure developers and model vendors can further explore scenarios where the link from the source node to the first replication node serves as the overall system bottleneck, so as to fully exploit the latency reduction brought by multicast communication.

\section{Design of the MultiWrite}
\label{sec:design}

After demonstrating practical real-world scenarios to verify the latency reduction capability of multicast, two remaining challenges still need to be addressed. The first is the explosive growth of traditional multicast groups, and the second is the lack of native multicast support within existing AI ecosystems. This section addresses these two issues one by one (Section \ref{sec:design_cha2} \& \ref{sec:design_cha3}), followed by an overview of the MultiWrite (Section \ref{sec:design_overview}).

\subsection{Handling Multicast Group Explosion}
\label{sec:design_cha2}
The core idea to solve this problem is to eliminate the reliance on multicast group establishment. Instead, the source node embeds metadata indicating destination node addresses into each outgoing packet. Upon receiving the packet, relay nodes parse the carried metadata and perform packet replication and forwarding according to preconfigured mapping rules.

\begin{figure*}[htbp]
	\centering
	\includegraphics[width=0.9\linewidth]{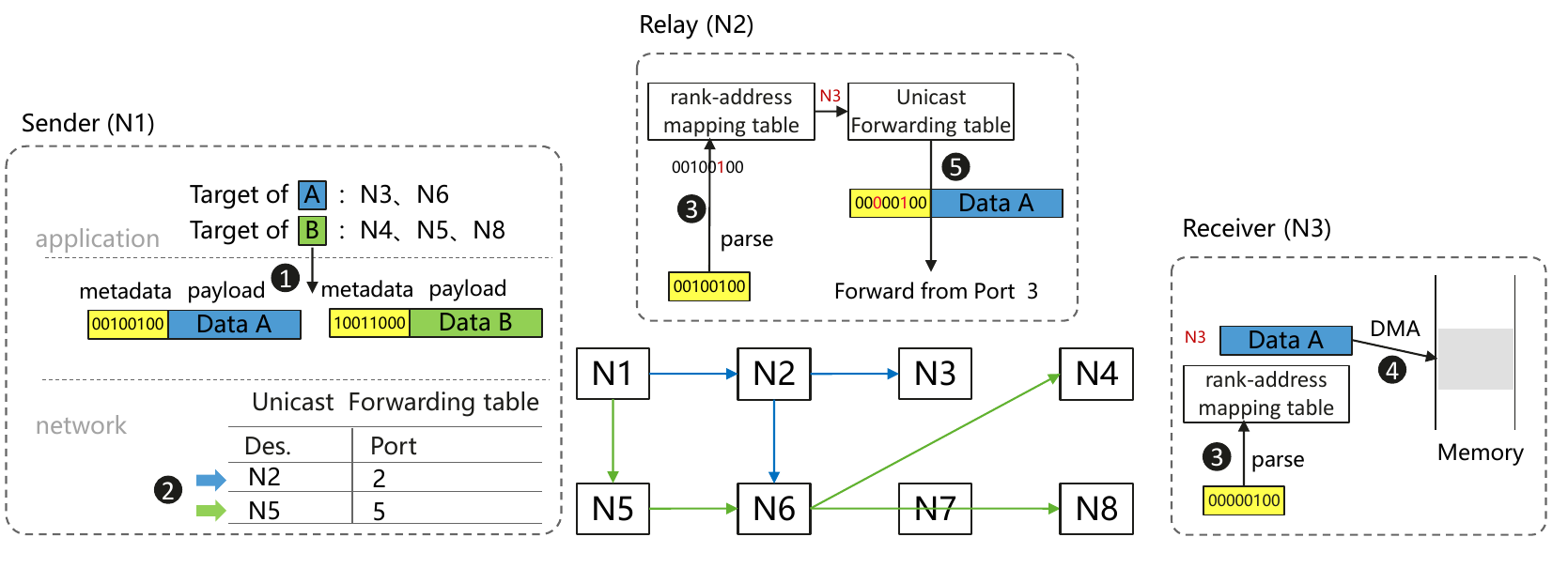}
	\caption{The sender embeds metadata indicating destination node addresses into each outgoing packet. Upon receiving the packet, relay nodes parse the carried metadata and perform packet replication and forwarding according to preconfigured mapping rules.}
	\label{fig:workflow}
\end{figure*}

Figure \ref{fig:workflow} illustrates the workflow.
\textcircled{1} The source node gets the destination address information of the message data from the application layer and encodes such information into metadata carried with the packet.
\textcircled{2} The source node forwards the packet to downstream nodes following the local forwarding table.
\textcircled{3} Upon receiving the packet, the downstream node parses the carried metadata.
\textcircled{4} If the node is one of the destinations, it delivers the payload to the target memory address.
\textcircled{5} Meanwhile, if the metadata contains other destinations, the node replicates the packet and forwards it to other destinations.

Several details of the above workflow are described as follows.

\textbf{Encoding scheme of metadata.}
A naive solution is to directly embed destination address, such as the EID in the UB protocol, into the packet header. This scheme suffers from two obvious drawbacks. It introduces non-negligible header overhead. Each EID occupies 128 bits. For a packet targeting eight destinations, a typical top-k setting in MOE scenarios, the metadata consumes 128 bytes. Furthermore, directly embedding raw addresses results in a variable-length header field when the set of destination nodes changes dynamically. Variable-length header fields severely degrade the efficiency of packet parsing and processing.

We use a bitmap encoding scheme based on NPU rank indices to specify destinations. We designate a fixed-size field in the packet header, for instance 64 bits, to hold a bitmap. All possible NPUs that may participate in the multicast are numbered starting from zero according to their ranks. Each bit in the bitmap indicates, with 1 or 0, whether the NPU with the corresponding index is a target of this packet.
This scheme requires a preconfigured software mapping table that maps physical addresses to rank IDs on each node. The table supports metadata encoding at the source node and metadata parsing at both relay and destination nodes.

\textbf{Local forwarding table.}
Since we establish no explicit multicast groups, our design avoids maintaining a dedicated multicast forwarding table as in traditional multicast schemes. Instead, we fully reuse the unicast forwarding table that each node already employs for regular unicast transmission. For a given packet, we look up the egress port for every destination node indicated in the bitmap metadata, compute the union of all distinct egress ports, and the number of unique egress ports determines how many copies the packet needs to be replicated. In cases where multiple egress ports exist for a single destination in the unicast forwarding table, we simply select one port following the same port-selection policy used in unicast routing.

\textbf{Replication and forwarding.}
In principle, packet replication and forwarding can be implemented at any layer above the network layer. However, to enable this design on existing off-the-shelf hardware without any firmware or hardware modifications, we implement the full replication logic at the application layer. Upon receiving a packet, the relay node first buffers the full payload, then parses the embedded metadata to determine the number of required packet copies. The process of computing the number of replicas follows the same logic as at the source node.

A critical detail in the forwarding procedure is the update of in-packet metadata at relay nodes. Packets sent out from different egress ports of the same relay node should carry modified metadata, so that downstream nodes will not perform redundant replication and forwarding, which would otherwise cause duplicate packet reception at destination nodes or routing loops. Notably, in simple topologies where routing loops are inherently impossible, and if end nodes implement packet deduplication at the receiver side, metadata updates can be omitted. Whether to perform metadata modification can be flexibly decided based on actual deployment scenarios.

\subsection{Bridging the Ecosystem Gap}
\label{sec:design_cha3}

With the elimination of dedicated multicast groups, all control-plane operations associated with multicast group management, including creation, deletion, and updating, become unnecessary. Only data-plane processing remains to be addressed. That is, how to coexist multicast logic with the existing unicast transmission mechanism. In conventional designs, such as the IB protocol, data-plane of multicast is enabled by attaching a queue pair to a multicast group, which operates independently from unicast communications. This approach relies on a tightly coupled design of both control plane and data plane. By abandoning explicit multicast groups, we propose a new solution to integrate multicast capabilities into current unicast-based communication stacks.

Our solution is to \textbf{semanticize multicast operations}. This design is motivated by the observation that nearly all collective communication operations in modern AI systems are implemented as semantic operations over existing transport protocols, among which the one-sided \texttt{write} operation is the most widely adopted. Since users and communication frameworks have been accustomed to the standard write semantic, we naturally define multicast as a new semantic named \texttt{MultiWrite}. It can be regarded as an enhanced extension of the write operation. A standard write instructs the receiver to write the payload into its local memory upon receiving the packet. In contrast, MultiWrite extends this behavior by replicating the data into multiple copies upon receiving a packet and delivering them to the memory of remote destination nodes.

We take the design of this semantic over the UB protocol as an example. In the UB protocol, semantic write is defined as a transaction operation, indicated by the \texttt{TAOpcode} field in the transaction header. For instance, opcode 0x3 represents a standard \texttt{write} operation, while 0x4 denotes \texttt{write\_with\_immediate}. Similarly, we introduce MultiWrite as a new transaction operation with a new \texttt{TAOpcode}. Packets associated with this new opcode are appended with metadata that indicates the complete set of multicast destinations.
During execution of a standard \texttt{write} transaction, the sender transmits this transaction to the receiver node. The receiver parses the write opcode and writes the payload to the designated local memory address.
MultiWrite follows the similar transaction workflow, except for the multicast-dedicated opcode. After parsing the opcode and metadata, the receiver performs remote memory writing for all target destinations. Specifically, if only one destination remains in the metadata, the node delivers the data via a standard write. If multiple destinations share the same next-hop relay, the current node initiates a nested child MultiWrite operation to forward the data to the next relay node.

\subsection{MultiWrite Semantics Overview}
\label{sec:design_overview}
\subsubsection {Semantic Definition and Abstraction}

MultiWrite is a unified one-sided memory semantic that extends the standard remote memory write operation to support multi-destination delivery. Unlike traditional multicast which is defined as a network-layer routing primitive, MultiWrite is defined at the application-transaction boundary, making it transport-agnostic and hardware-independent.

The core abstraction of MultiWrite can be formally stated as:
Given a source node $S$, a set of destination-memory pairs $M={(D_1, B_1), (D_2, B_2), ..., (D_n, B_n)}$ where $D_i$ is the $i$-th destination node and $B_i$ is the corresponding target memory buffer at $D_i$, and a source memory buffer $B_S$ at node $S$, MultiWrite($S$, $M$, $B_S$) atomically writes the content of buffer $B_S$ to the corresponding memory locations $B_i$ at all destination nodes $D_i$.

\subsubsection {Programming Interface}
From the perspective of upper-layer frameworks and applications, MultiWrite presents a single-function interface identical in usage to standard write operations. This design ensures minimal adoption barrier and seamless integration with existing communication libraries.
This interface is intentionally designed to mirror the standard \texttt{write()} interface. The only difference is that it accepts an array of remote buffer addresses instead of a single one. All complexity of in-network replication, routing, and metadata management is completely hidden from the caller.

\subsubsection {Recursive Execution Model}
MultiWrite follows a purely recursive execution model that requires no global coordination or preconfiguration. The execution can be described by three simple rules that apply uniformly to all nodes in the network.
\textcircled{1} When a node receives a MultiWrite request targeting a set of destination-memory pairs $M$:
\textcircled{2} If $|M|==1$, the node degenerates the operation into a standard write to the single destination and its corresponding target buffer.
\textcircled{3} If $|M|>1$, the node partitions $M$ into $l$ subsets $M_1, ..., M_l$ where each subset $M_i$ corresponds to a unique next-hop relay node. The node then initiates $l$ separate MultiWrite operations, each targeting one subset $M_i$.

This recursive model is the key to MultiWrite's simplicity and scalability. It ensures that the same logic is executed at every node in the path, from the source to the final destinations, without any special-case handling.

\subsubsection {Key Semantic Properties}
MultiWrite guarantees the following semantic properties that are critical for AI collective communication:

\noindent \textbf{Atomicity per destination.} The write to each individual destination's target buffer is atomic, matching the guarantee of standard write operations.

\noindent \textbf{No ordering guarantees between destinations.}
Writes to different destination buffers may complete in any order, which is consistent with the requirements of most AI collective operations.

\noindent \textbf{End-to-end reliability.}
MultiWrite inherits unicast reliability guarantees from the underlying transport protocol.

\noindent \textbf{Statelessness.}
No state is maintained in the network for MultiWrite operations. All necessary destination-memory mapping information is carried in the packets themselves.

These properties make MultiWrite a drop-in replacement for existing multi-destination communication patterns in AI training frameworks, while providing significant performance improvements.

\section{Implementation}

We implement the proposed MultiWrite semantic as a pure software extension on top of the UB protocol stack (Section \ref{sec: imp1}). Based on this semantic, we further develop MultiWrite-based collective communication operations for AI cluster workloads (Section \ref{sec: imp2}).

\subsection{Implementation of MultiWrite}
\label{sec: imp1}
In the UB ecosystem, UMDK (Unified Memory Development Kit) serves as the counterpart of OFED (OpenFabrics Enterprise Distribution) in the RDMA ecosystem, providing a complete software stack including kernel drivers (e.g. \texttt{ubcore.ko}), user-space libraries (e.g., \texttt{liburma.so}), and management tools (e.g., \texttt{urma\_perftest}) to enable UB protocol-based communication. UMDK has been open-sourced along with the opening of the UB protocol\cite{umdk_gitcode}.

\begin{figure}[htbp]
	\centering
	\includegraphics[width=0.8\linewidth]{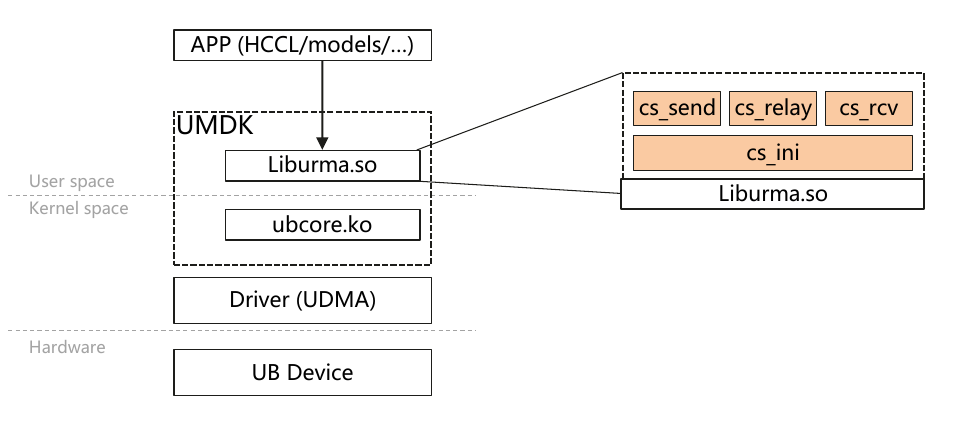}
	\caption{The position of MultiWrite modules in the system stack, where colored components denote the newly designed modules in this work.}
	\label{fig:modules}
\end{figure}

MultiWrite is built as an extension on top of the existing \texttt{liburma.so} library, as shown in \ref{fig:modules}. It consists of four core modular components that jointly realize the complete semantic.
\begin{itemize}
	\item The \texttt{cs\_ini} module runs on every node that uses this semantic and performs essential initialization operations, including loading topology and node configuration information and launching the daemon relay threads.
	\item The \texttt{cs\_relay} module acts as a daemon thread created by the initialization module, with a lifecycle consistent with the user application. It is responsible for packet coping and forwarding.
	\item The \texttt{cs\_send} module handles task submission at the source node, including task decomposition and generation of corresponding WQEs for transmission.
	\item The \texttt{cs\_rcv} module manages packet reception, reassembly, and delivery of the final data buffer addresses to the user application.
\end{itemize}

\subsection{Implementation of MultiWrite-based collective communication operators}
\label{sec: imp2}

To demonstrate the practical utility of the proposed MultiWrite semantic, we take the MultiWrite-based AllGather operation described in Section \ref{sec:8pfullmesh} as a representative example. We first outline the baseline workflow of traditional AllGather in a full-mesh topology, then present our optimized implementation leveraging MultiWrite to address the inter-TP link underutilization problem.

In a full-mesh topology, the traditional AllGather operation (TP=4) follows the following workflow.
\textcircled{0} Establish a communication domain. All nodes establish pairwise connections and exchange critical information including buffer addresses, sizes, and node identifiers.
\textcircled{1} Data preparation. Each node loads its local message data into the pre-allocated send buffer.
\textcircled{2} Concurrent transmission. Every node initiates three non-blocking asynchronous unicast sends simultaneously.
\textcircled{3} Reception completion check. Nodes continuously poll the corresponding fields in their local receive buffers to verify that data from all remote nodes has been successfully written.
\textcircled{4} Global synchronization. A global barrier operation is performed to confirm that all four nodes have received all required data chunks.

As analyzed in Section \ref{sec:8pfullmesh}, this traditional approach does not leave inter-TP domain links. To resolve this issue, we deploy the MultiWrite modules described earlier on each node and modify the AllGather workflow as follows:
\textcircled{0} Extended communication domain. We establish a unified communication domain spanning all eight nodes, exchanging complete topology information and buffer configurations to enable cross-domain transmission.
\textcircled{1} Data partitioning and preparation: Each node splits its local data into two segments. One segment is transmitted via intra-TP direct links, while the other is sent through inter-TP domain links. The split ratio is dynamically calculated based on the measured bandwidth of both link types, ensuring that data from both paths arrives at each destination node simultaneously to minimize overall latency.
\textcircled{2} Hybrid transmission. Each node concurrently initiates three standard unicast writes over intra-TP direct links, plus one cross-TP MultiWrite operation that delivers the second data segment to all four nodes.
\textcircled{3} \& \textcircled{4} The reception check and global barrier steps remain identical to the traditional implementation, requiring no modifications to existing upper-layer collective logic.

\section{Evaluation}
\label{sec:eva}

We evaluate MultiWrite with experiments to answer the following questions:

We conduct comprehensive experiments to evaluate the practical performance of the proposed MultiWrite semantics in real-world commercial environments and answer the following questions:
\begin{itemize}
\item Can collective communication operators enhanced by MultiWrite, such as AllGather and AlltoAll, reduce end-to-end latency in real commercial devices? (Section \ref{sec:exp_latency})
\item What is the capability boundary of MultiWrite? Do its performance gains remain consistent across different message sizes? (Section \ref{sec:exp_scale})
\item What overhead does MultiWrite introduce when achieving latency improvement? (Section \ref{sec:exp_cost})
\end{itemize}

\subsection {Experiment setup}
\label{sec:exp_setup}

\textbf{Cluster.}
We conduct all experiments on a NPU clusters of two servers. Each server is equipped with four 48-core Kunpeng-920 CPUs and eight 910B4 NPUs. Within each server, every NPU is interconnected in a full-mesh topology via the Huawei Cache Coherence System (HCCS). Each HCCS link provides a physical bandwidth of 56 GB/s. Meanwhile, each NPU is integrated with a RoCE network interface card. Cross-server NPU-to-NPU communication relies on these RoCE NICs, which are deployed within a CLOS-based network fabric with a per-port egress bandwidth of 200 Gbps.

The Ascend 910B NPU chip has been commercially available and stably deployed in production for over four years, serving as a mature and cost-effective computing chip. The 910B series includes multiple models with varying HBM capacities. In this work, we select the configuration with the smallest memory specification, namely 910B4, which is equipped with only 32 GB HBM. By conducting evaluations on this prevalent yet memory-constrained model, we demonstrate the general applicability and practicality of our proposed solution.

\noindent \textbf{Workload.}
We evaluate two collective communication operations.
\begin{itemize}
\item AllGather: We partition the eight NPUs inside a single server into two tensor parallel groups of size 4, following the configuration described in Section \ref{sec:8pfullmesh}. The end-to-end latency of the AllGather operator is measured using an industry-standard long-duration stress test.

\item AlltoAll: For AlltoAll, we target the dispatch phase in MoE workloads. The evaluation is conducted on 16 NPUs with 64 deployed experts, where each token selects the top-8 experts for computation. To ensure a fair evaluation, expert load balancing is enabled throughout all experiments.
\end{itemize}

\noindent \textbf{Baselines.}
We choose the officially released and production-ready \texttt{HcclAllGather}\cite{hcclallgather} and \texttt{HcclAlltoAllV}\cite{hcclalltoallv} in HCCL as the baselines for our experiments. Notably, we integrate several internal optimization tricks into the baseline, which are not enabled in official commercial releases due to additional usage complexity. Since these tricks further strengthen rather than degrade the baseline performance, we omit their detailed descriptions in this paper.
In addition to the standard baseline, we implement an optimized AllGather design, which leverages unicast multipath leveraging, as described in Section \ref{sec:8pfullmesh}.

\textbf{Metrics.}
We measure the end-to-end latency of an AllGather or AlltoAll operation, which is calculated from the moment the operation API is called until it returns with the complete output data in the caller's buffer. The reported global end-to-end latency is the maximum across all participating NPUs. To prevent cold-start overheads from affecting the evaluation, we always execute several warm-up iterations before timing the measured iterations. We will present the per-iteration latency across a long-duration stress test. When performing cross-scheme comparisons, we use the average latency of the measured iterations as the final result. In addition to the latency, we further evaluate the overhead of our approach in Section \ref{sec:exp_cost}.

We focus purely on collective communication performance instead of model-level results for two reasons. First, this study must follow confidentiality constraints. Second, collective communication typically occupies 10\% to 30\% of total runtime, and can reach nearly 50\% in contrived scenarios to inflate overall gains. We consider such selective evaluation misleading. Based on our communication-level results, readers can estimate end-to-end improvements according to the communication ratio of their own model workloads.

\subsection {Latency Reduction}
\label{sec:exp_latency}

\begin{figure}[htbp]
\centering
\includegraphics[width=1.0\linewidth]{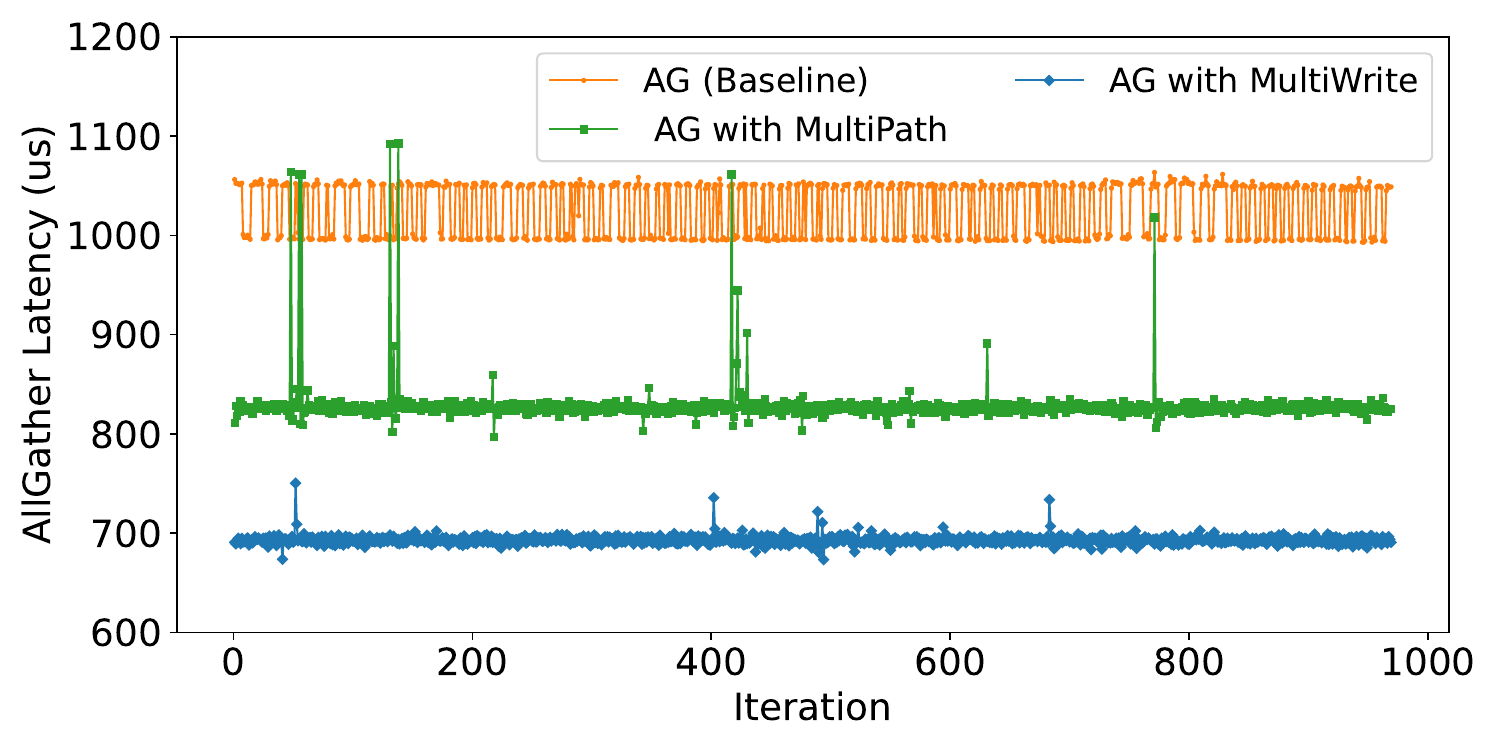}
\caption{AllGather latency corresponding to three schemes. The results are collected from nearly 1000 iterations of online stress tests for each AllGather communication operator. The AllGather operator built upon MultiWrite achieves the lowest and most stable latency.}
\label{fig:allgather_timeseries}
\end{figure}

Figure \ref{fig:allgather_timeseries} presents the end-to-end latency of AllGather operations under stress tests. In this experiment, each NPU transmits 16 MB of data, a typical message size for AllGather workloads. The orange curve denotes the baseline, the green curve represents the unicast multipath optimized AllGather, and the blue curve is AllGather built upon MultiWrite. In the baseline, data is transferred only over direct intra-server links. The two optimized schemes additionally leverage inter-TP-domain links. The unicast multipath scheme duplicates data and transmits redundant copies, whereas the MultiWrite scheme completes the transfer without redundancy.

The MultiWrite-based AllGather consistently achieves the lowest latency, reducing it by approximately 30\% compared with the baseline and by 17\% compared with the unicast multipath scheme. The unicast multipath scheme exhibits noticeable latency variance and sometimes performs even worse than the baseline, because it initiates multiple concurrent flows on inter-TP-domain links, which are more susceptible to mutual interference and resource contention.

We also conduct the same stress test for AlltoAll latency. Due to space limitations, these results are presented together in the next subsection.

\subsection {Sensitivity to Message Scale}
\label{sec:exp_scale}
The latency reduction reported above is obtained at a fixed message size. To more fully characterize the capability boundary of MultiWrite, we evaluate both AllGather and AlltoAll across the message sizes commonly encountered in practice.

\begin{figure}[htbp]
\centering
\includegraphics[width=0.8\linewidth]{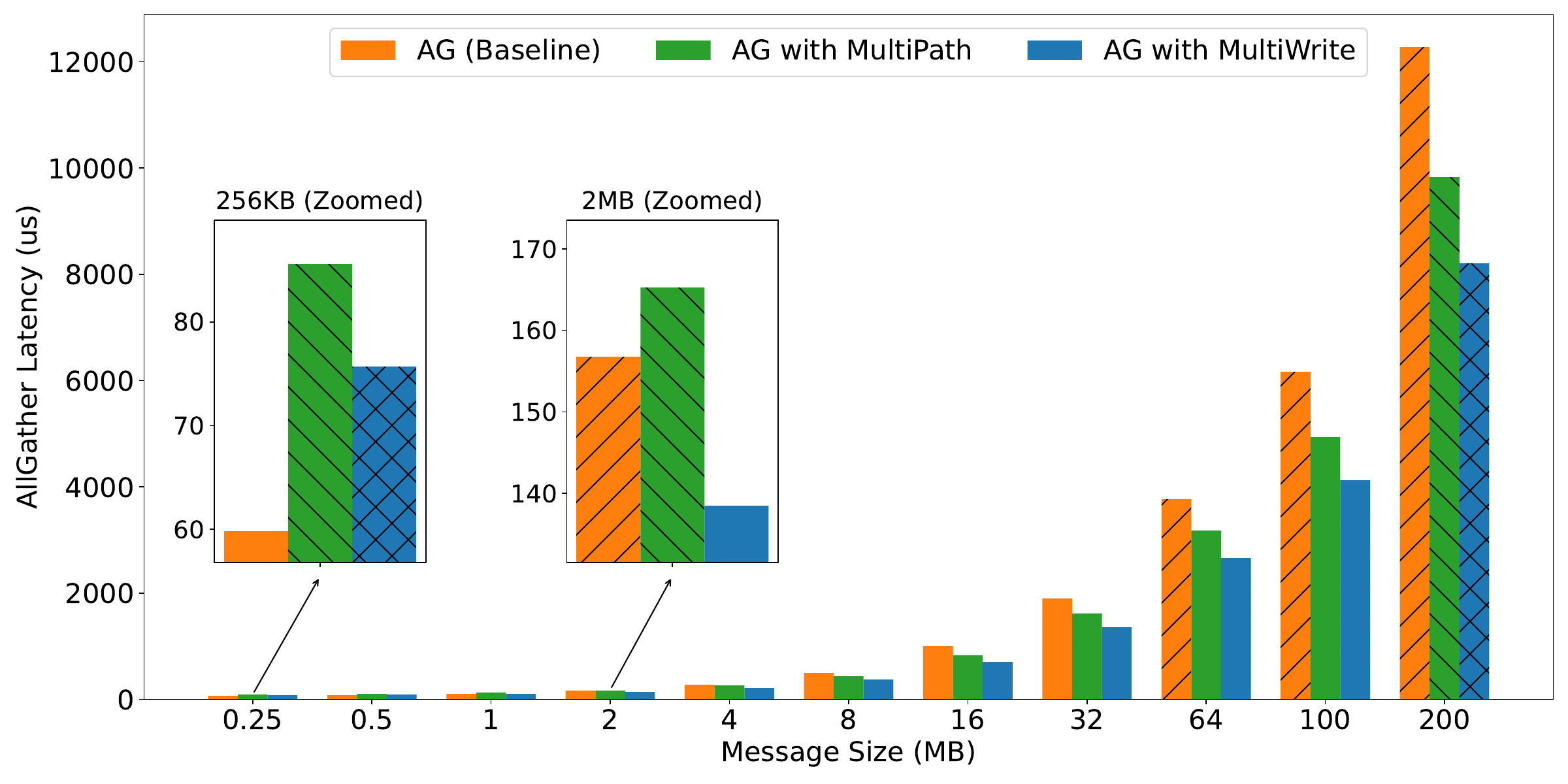}
\caption{End-to-end latency of AllGather operators under different message sizes.}
\label{fig:allgather_datasize}
\end{figure}

\textbf{AllGather.}
Empirically, per-rank message sizes for AllGather range from a few megabytes to several hundred megabytes. We therefore test message sizes from 256 KB to 200 MB. Figure \ref{fig:allgather_datasize} presents the results, with yellow and blue bars representing the baseline and the MultiWrite-based AllGather, respectively. At large message sizes, the latency gains of MultiWrite-based AllGather closely match the results shown in Figure \ref{fig:allgather_timeseries}. However, when the message size is small, such as 256 KB, MultiWrite-based AllGather incurs higher latency than the baseline, with the crossover occurring around 2 MB. The reason is that both unicast multi-path and MultiWrite optimizations essentially exploit inter-TP-domain links to increase effective bandwidth, which requires extra data processing such as splitting. When the per-rank message is small, it is already difficult to saturate a single link. Multi-path transmission then brings no benefit and instead adds extra processing overhead. Fortunately, in most AllGather transmission, the data size per-rank is no less than 1 MB, which falls within the region where MultiWrite is advantageous.

The practical data size per transmission rarely exceeds 200 MB in real-world deployments. For workloads requiring larger data transfer, collective communication frameworks will automatically split oversized messages into multiple chunks below 200 MB, due to considerations such as communication buffer pre-allocation. Therefore, we set 200 MB as the upper bound in our evaluation.

\begin{figure}[htbp]
\centering
\subcaptionbox{}{
\includegraphics[width=0.45\linewidth]{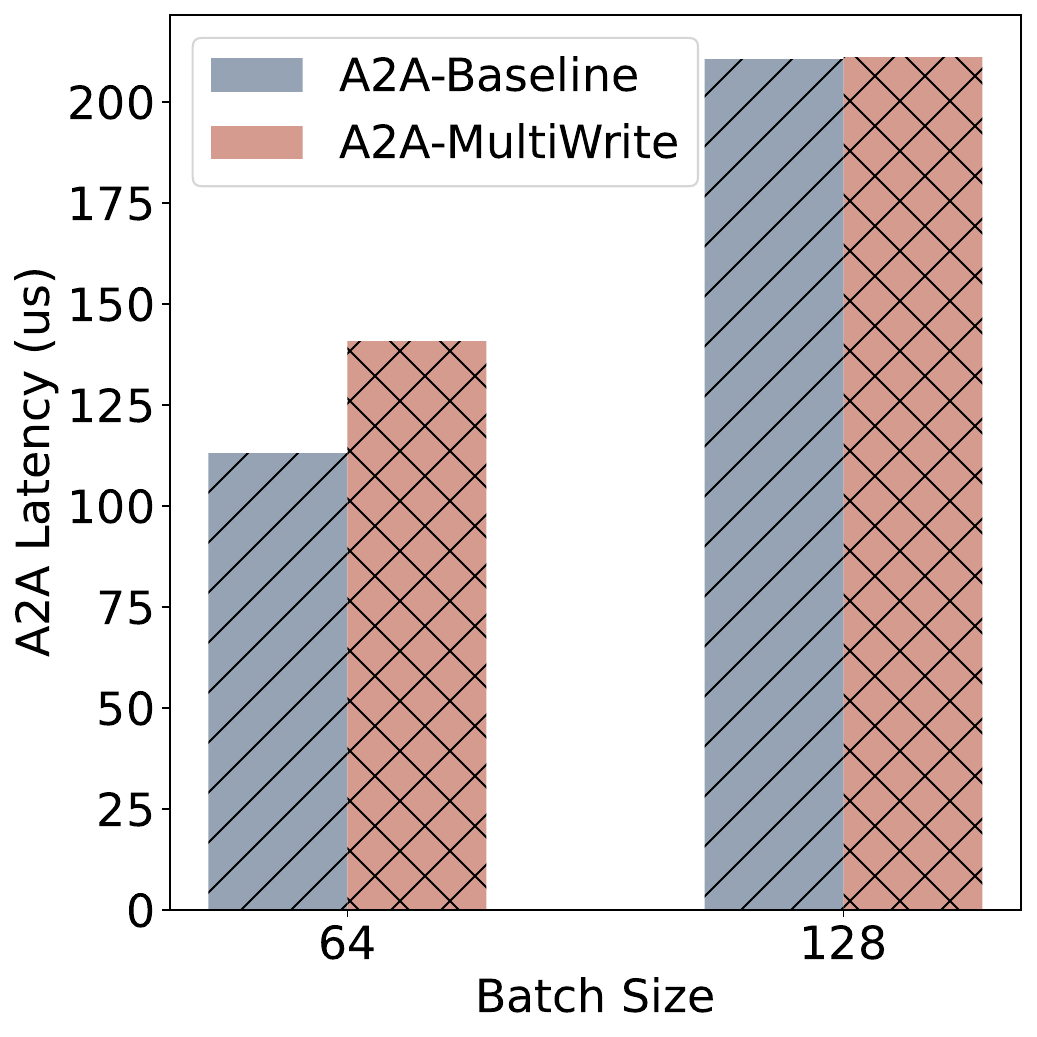}
}%
\hfill
\subcaptionbox{}{
\includegraphics[width=0.45\linewidth]{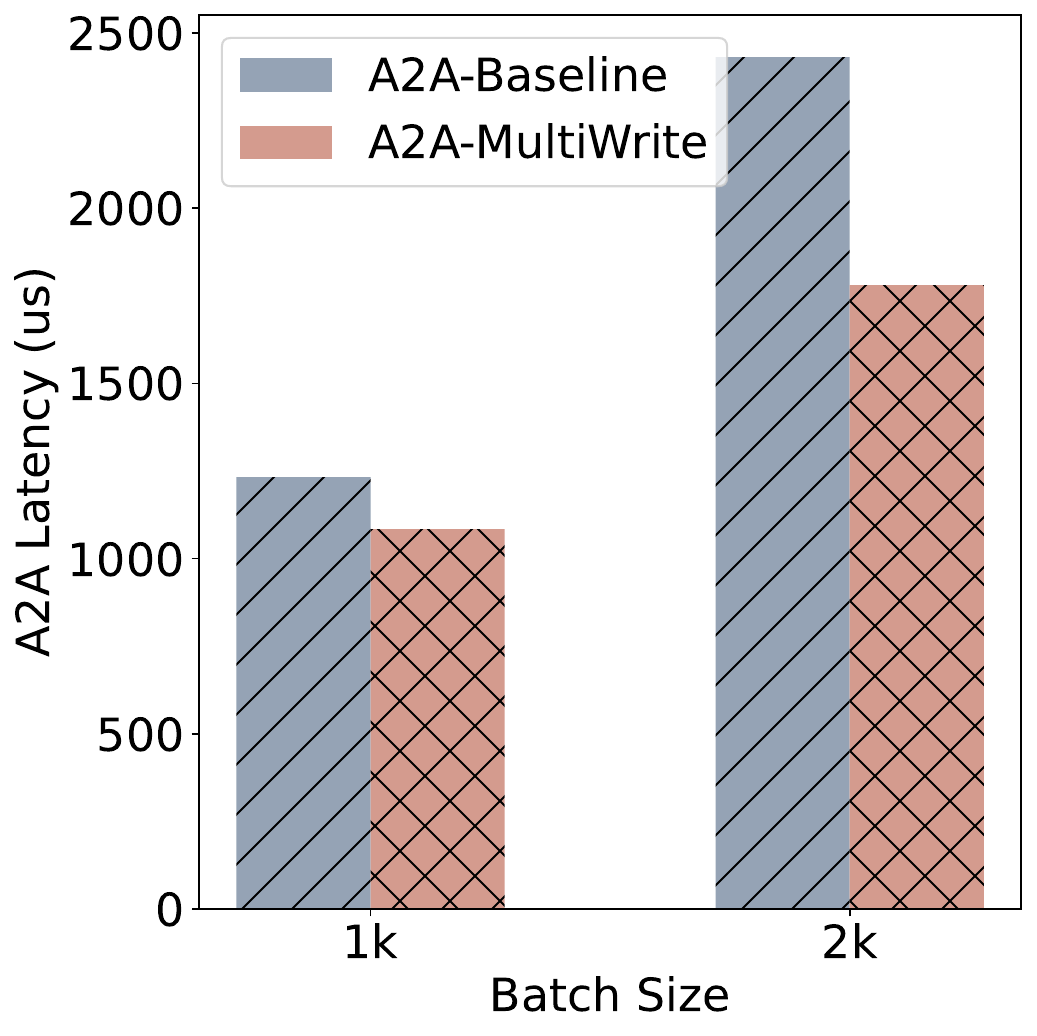}
}%
\caption{End-to-end latency of AlltoAll operators under different batch sizes. (a) Decode phase with typical sizes of 64 and 128; (b) Prefill phase with typical sizes of 1k and 2k. The MultiWrite-based AlltoAll operator achieves more significant performance gains under large batch sizes.}
\label{fig:a2a-all}
\end{figure}

\textbf{AlltoAll.}
AlltoAll dispatch operators are primarily adopted in MoE workloads, where the separation of prefill phase and decode phase has become a de facto standard. To balance throughput and latency, the batch size used in these two phases are different. The batch size here refers to the number of tokens processed by experts in a single batch. The decode phase typically has a small batch size of dozens of tokens, while the prefill phase usually operates at a scale of several thousand tokens. We evaluate AlltoAll performance for both phases.

Figure \ref{fig:a2a-all}-(a) presents the AlltoAll results with typical decode phase batch sizes of 64 and 128. At a batch size of 64, the MultiWrite-based AlltoAll exhibits higher latency than the baseline. The two schemes achieve nearly identical latency when the batch size rises to 128. Figure \ref{fig:a2a-all}-(b) illustrates the results for prefill phase batch sizes of 1k and 2k. In these two batch sizes, the MultiWrite-based AlltoAll reduces latency by 12\% and 27\%, respectively, compared with the baseline.

\begin{table}[htbp]
\centering
\caption{Cross-server Transfer Latency: With and Without Redundant Data Replicas}
\label{tab:cross-server}
\begin{tabular}{ccccc}
	\toprule
	BatchSize (tokens) & 64 & 128 & 1k & 2k \\
	\midrule
	w/ redundant & 112.90 & 210.53 & 1231.18 & 2429.72 \\
	w/o redundant & 43.77 & 66.63 & 320.52 & 622.10 \\
	\midrule
	$\Delta$ (us)  & 69.13 & 143.90 & 910.66 & 1807.62 \\
	\bottomrule
\end{tabular}
\end{table}

The reason is that, although eliminating redundant cross-server transfers does reduce the inter-server transmission time, it introduces extra latency from token replication at the same-index NPU and subsequent intra-server forwarding. When the batch size is small, the inter-server savings are insufficient to offset the additional intra-server processing and transfer costs. Table \ref{tab:cross-server} confirms this observation. It reports the cross-server transfer times with and without redundant data. The MultiWrite-based AlltoAll removes that redundancy, but at small batch sizes the volume of redundant traffic is low, so the time saved is modest.

\subsection {Cost Analysis}
\label{sec:exp_cost}
The overhead of MultiWrite comes from three aspects.

\textbf{Metadata overhead.}
As discussed in Section \ref{sec:design}, a core feature of MultiWrite is the removal of multicast groups, where destination information is instead embedded within each packet. In our design, a 64-bit bitmap encodes all NPU ranks in the communication domain, with each bit indicating whether the corresponding rank is a target node. This metadata is transmitted via the immediate field of the \texttt{write\_with\_immediate} operation. Both UB and RDMA protocols natively reserve this field. Consequently, this design introduces no additional packet header overhead compared with existing schemes.

When the communication domain scale exceeds 64 NPUs, such as a 128-NPU deployment, two solutions are available. The first leverages reserved fields predefined in protocols, which are universally for existing protocols. The second embeds metadata into the packet payload rather than the header. Although this approach slightly reduces effective payload capacity, the impact is marginal. Taking the UB protocol as an example, for a communication domain with 1024 NPUs, the metadata occupies only 3.13\% (1024 bit/4096 byte) of the total payload size. Overall, the overhead introduced by metadata remains negligible and practically acceptable.

\textbf{CPU usage.}
Relay NPUs need to parse metadata, and perform packet copying and forwarding, which introduces extra CPU overhead. Since our scheme is deployed on devices, the CPU here refers to the CPU on the device, not the host. In our implementation, it refers to AICPU on Ascend chips. We use \texttt{npu-smi}, a dedicated tool for Ascend, to compare AICPU usages between the baseline and MultiWrite-based AllGather under identical transmission configurations.

\begin{figure}[htbp]
\centering
\includegraphics[width=0.9\linewidth]{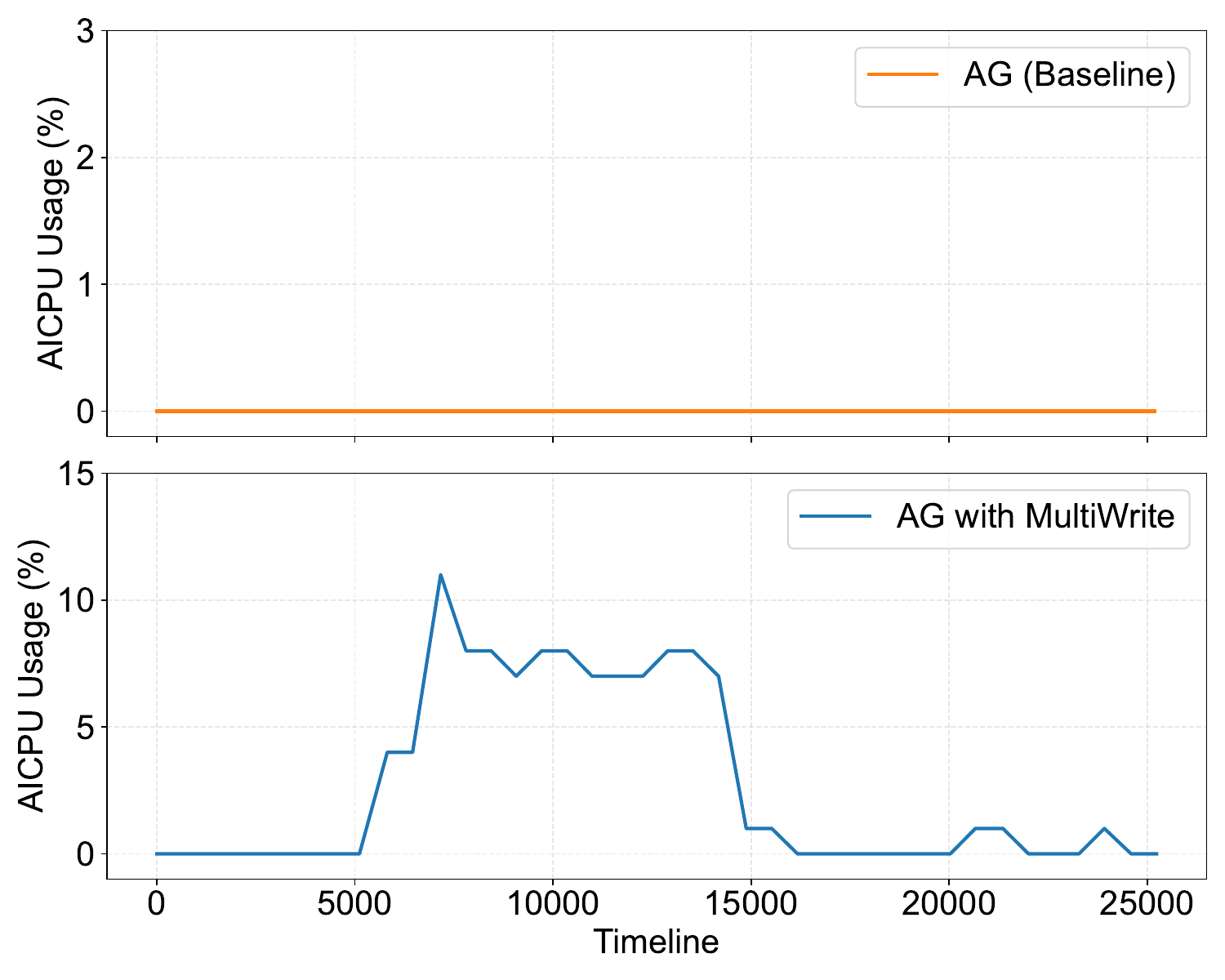}
\caption{AICPU usage of the baseline and MultiWrite-based AllGather. The profiling is conducted with a per-rank message size of 200 MB.}
\label{fig:cost}
\end{figure}

Figure \ref{fig:cost} shows the result. MultiWrite brings a moderate increase in AICPU usage compared with the baseline. It is worth noting that the computing capability of AICPU is relatively limited. In practical deployment, existing workloads generally bypass AICPU for high-performance data-plane operations, and rely instead on AICube and AIVector within AICORE. This is why the AICPU usage corresponding to the baseline is almost zero. As a result, AICPU remains idle in most scenarios. Although MultiWrite raises AICPU usage, this extra load has negligible impact on co-running services.

\textbf{Memory usage.}
Relay NPUs needs device memory to receive data from the source node, which is then replicated and forwarded. In our implementation, this relay buffer is exactly the same as the receive buffer. For instance, when NPU1 sends packets to NPU9 for relay, all incoming data to NPU9, including both forwarded traffic and data destined for itself, is first buffered before processing. Even without MultiWrite, current collective communication frameworks already reserve dedicated buffers for rank-to-rank data reception. MultiWrite simply reuses these pre-allocated receive buffers for relay tasks. Therefore, compared with the baseline, MultiWrite introduces no significant additional memory usage.

\section{Related Work}
\label{sec:relatedwork}

\textbf{Hardware-native multicast.}
Hardware-native multicast has long been considered the most efficient way to implement one-to-many communication in high-performance networks. InfiniBand provides native multicast support through unreliable datagram (UD) queue pairs attached to multicast groups\cite{ibta2024infiniband}. RoCEv2 also extends this multicast capability to standard Ethernet fabrics, enabling RDMA multicast over commodity Ethernet switches.

However, hardware multicast suffers from several critical limitations. First, it requires pre-establishment and management of multicast groups, which introduces significant control-plane overhead and scalability challenges\cite{mamidala2005efficient}.
Second, native InfiniBand multicast is inherently unreliable\cite{ibta2024infiniband}. This requires applications to implement complex retransmission and ordering mechanisms at the upper layer\cite{khalilov2024network, liu2004fast, hoefler2007practically, liu2014ibrmp}.
Third, hardware multicast operates independently from unicast traffic, requiring separate queue pairs, buffer management, and congestion control mechanisms\cite{liu2014ibrmp, liu2004fast}.
Finally, hardware multicast is tightly coupled with the underlying network infrastructure. This makes it difficult to deploy on existing commodity hardware.

\noindent \textbf{Software multicast.}
Given that this work targets AI training and inference clusters, we first focus on software multicast designs over RDMA.

Several recent efforts have explored software multicast over RDMA networks to overcome the deployment constraints of hardware multicast. Huang \textit{et al.} proposed a software reliable multicast that improves replication performance in RDMA-based distributed systems\cite{huang2023mc}, demonstrating its viability for data replication and consensus without specialized hardware. Li \textit{et al.} presented Cepheus\cite{li2024cepheus}, which reuses RoCE as its transport layer and provides a RoCE-capable multicast primitive via in-network assistance. A key commonality between these two works is that both leverage the Reliable Connection (RC) transport mode of RoCE and exploit the forwarding capabilities of intermediate nodes, thereby making multicast behavior compatible with existing reliable unicast one-to-one transmissions. In contrast, Khalilov \textit{et al.} takes a different approach \cite{khalilov2024network}. This work builds upon InfiniBand hardware multicast, which inherently uses unreliable datagram transport, and adds software-level reliability guarantees at the end hosts to achieve lossless collective operations.

A key limitation of all these software multicast approaches is that they are all based on fixed multicast groups, which cannot support scenarios where destination nodes change on a per-token basis, such as AlltoAll dispatch in AI training workloads.

Besides the RDMA protocol, other interconnection technologies either remain closed-source, such as NVLink\cite{nvidianvlink}, making their design details unavailable to us, or provide no public documentation regarding multicast support, such as UALINK\cite{ualinkspec}, UEC\cite{ultraethernetspec}, SUE\cite{ocpsuespec}, and ESUN\cite{ocpesunintro}. Notably, a scheme in traditional IP networks shares similar design philosophy with our work, \textit{i.e.} BIER (Bit Indexed Explicit Replication)\cite{rfc8279, rfc8296, rfc9793}. The core idea of BIER is to embed a bitstring in the packet header, where each bit corresponds to a Bit-Forwarding Router (BFR) in the network. Routers parse the bitstring to directly determine whether to replicate and forward the packet.
Despite differences in concrete metadata encoding formats and their definitions, this mechanism is similar to our approach in addressing the multicast group explosion problem. In fact, carrying destination information inline with data packets is a classic and natural design design. The critical limitation of BIER is that it is implemented as a tightly coupled full-stack technology built on top of IP, while IP is no longer the preferred choice for high-speed, low-latency transmission in modern AI clusters, especially within scale-up domains.

\noindent \textbf{Collective Communication Optimizations for AI Clusters.}
Collective communication operations often become a major performance bottleneck in large-scale AI training workloads.
Industrial and academic communities have developed highly optimized collective communication libraries to address this challenge. NVIDIA's NCCL serves as the prevailing standard for GPU collective communication in AI clusters. Microsoft's MSCCL++ \cite{microsoft2022msccl} extends NCCL with a domain-specific language for custom collective algorithms, enabling drop-in acceleration of existing PyTorch applications.
Huawei's HCCL is the widely adopted native collective communication library for Ascend NPU-based AI training clusters.

These libraries primarily focus on optimizing communication algorithms within a given network topology, but they are fundamentally limited by the underlying transport primitives. Traditional AllGather implementations rely on unicast communication. While recent work has explored topology-aware collective algorithms\cite{cai2021synthesizing, shah2023taccl}, they still operate at the level of individual unicast operations.

Our work differs from existing collective communication optimizations in that we introduce a new semantics that natively supports multi-destination memory writes. This allows us to implement collective operations such as AllGather with significantly higher bandwidth utilization, especially in hierarchical topologies where inter-domain links are otherwise underutilized. Unlike previous approaches, MultiWrite is implemented entirely in software on top of existing memory semantics, requiring no modifications to hardware or transport protocols.
\section{Conclusion}

In this work, we target the critical challenges of multicast communication in AI training and inference clusters. Conventional multicast mechanisms suffer from three core drawbacks: insufficient recognition of its latency reduction value, severe multicast group explosion, and poor compatibility with modern AI computing ecosystems.
We present MultiWrite, a semantics that extends generic memory semantics to natively support efficient multi-destination data delivery. This design fully exploits the latency benefits of one-to-many communication, and achieves natural compatibility with AI cluster communication scenarios. Long-term commercial stress tests demonstrate that the proposed MultiWrite-based operators achieve up to 33\% latency reduction on deployed devices.


\bibliographystyle{ACM-Reference-Format}
\bibliography{ref/references}

\end{document}